\documentstyle[12pt,psfig]{article}
\begin{document}
TPI-MINN-01-36 $\;\;$
          UMN-TH-2020-01 $\;\;$
\newcommand{\nc}{\newcommand}
\def\bo{\mathrel{\raise.3ex\hbox{$>$\kern-.75em\lower1ex\hbox{$\sim$}}}}
\def\me{\mathrel{\raise.3ex\hbox{$<$\kern-.75em\lower1ex\hbox{$\sim$}}}}
\nc{\gone}{\bar g_{\pi NN}^{(1)}}
\nc{\gzero}{\bar g_{\pi NN}^{(0)}}
\nc{\al}{\alpha}
\nc{\ga}{\gamma}
\nc{\de}{\delta}
\nc{\ep}{\epsilon}
\nc{\ze}{\zeta}
\nc{\et}{\eta}
\renewcommand{\th}{\theta}
\nc{\Th}{\Theta}
\nc{\ka}{\kappa}
\nc{\la}{\lambda}
\nc{\rh}{\rho}
\nc{\si}{\sigma}
\nc{\ta}{\tau}
\nc{\up}{\upsilon}
\nc{\ph}{\phi}
\nc{\ch}{\chi}
\nc{\ps}{\psi}
\nc{\om}{\omega}
\nc{\Ga}{\Gamma}
\nc{\De}{\Delta}
\nc{\La}{\Lambda}
\nc{\Si}{\Sigma}
\nc{\Up}{\Upsilon}
\nc{\Ph}{\Phi}
\nc{\Ps}{\Psi}
\nc{\Om}{\Omega}
\nc{\ptl}{\partial}
\nc{\del}{\nabla}
\nc{\be}{\begin{eqnarray}}
\nc{\ee}{\end{eqnarray}}
\nc{\ov}{\overline}
\nc{\gsl}{\not\!}
\newcommand{\s}{\mbox{$\sigma$}}
\newcommand{\bi}[1]{\bibitem{#1}}
\newcommand{\fr}[2]{\frac{#1}{#2}}
\newcommand{\gm}{\mbox{$\gamma_{\mu}$}}
\newcommand{\gn}{\mbox{$\gamma_{\nu}$}}
\newcommand{\Le}{\mbox{$\fr{1+\gamma_5}{2}$}}
\newcommand{\R}{\mbox{$\fr{1-\gamma_5}{2}$}}
\newcommand{\GD}{\mbox{$\tilde{G}$}}
\newcommand{\gf}{\mbox{$\gamma_{5}$}}
\newcommand{\Ima}{\mbox{Im}}
\newcommand{\Rea}{\mbox{Re}}
\newcommand{\Tr}{\mbox{Tr}}
\newcommand{\cp}{\;\;\slash{\!\!\!\!\!\!\rm CP}}
\newcommand{\qq}{\langle \ov{q}q\rangle}
\newcommand{\qqg}{\langle \ov{q}\ga_5q\rangle}
\newcommand{\psl}{\slash{\!\!\!p}}
\newcommand{\xsl}{\slash{\!\!\!x}}
\newcommand{\uu}{\ov{u}u}
\newcommand{\dd}{\ov{d}d}
\newcommand{\uGu}{\ov{u}g_sG\si u}
\newcommand{\dGd}{\ov{d}g_sG\si d}
\nc{\go}{\bar g_{\pi NN}^{(1)}}
\nc{\gz}{\bar g_{\pi NN}^{(0)}}


\begin{center}

{\bf Best Values for the CP-odd Meson-Nucleon Couplings
from Supersymmetry
}

\vspace{1cm}

Maxim Pospelov

\vspace{1cm}
 {\it Theoretical Physics Institute, School of Physics and Astronomy \\
         University of Minnesota, 116 Church St., Minneapolis, MN 55455,
         USA}\\and \\ {\it
         Centre for Theoretical Physics,
                                 CPES,
                                 University of Sussex,
                                 Brighton BN1 9QJ,
                                 United Kingdom}\footnote{Present address}\end{center}



\vspace{1cm}

\begin{abstract}
In the supersymmetric models, the dominant
sources of the hadronic flavor-diagonal CP violation at low energy
are the theta term and the chromoelectric dipole moments of quarks.
Using QCD sum rules, we estimate the preferred range and the best values
for the CP-odd meson-nucleon coupling constants induced by these
operators. When the theta term is removed
by the axion mechanism, the size
of the most important isospin triplet pion-nucleon coupling is estimated to be
$\bar g_{\pi NN}^{(1)} = 2\times 10^{-12}(\tilde d_u - \tilde d_d)$, where
chromoelectric dipole moments are given in units of $
10^{-26}{\rm cm}$.
\end{abstract}

\newpage 

The search for CP violation in flavor-conserving processes
is of paramount scientific importance. The suppression of
CP-violating effects induced by the complex phase of the
Kobayashi-Maskawa matrix allows the use of electric
dipole moments (EDMs) of neutrons or heavy atoms as well as
T-odd asymmetries in the decays and scattering of baryons as powerful
tools for probing new physics beyond the Standard Model.

The wide separation between the energy scale
of ``new physics'' (superpartners,
technicolor, etc.) and the characteristic momenta of particles
in non-accelerator experiments permits consideration of only the first
few terms in the effective CP-odd Lagrangian. In the minimal supersymmetric
models only the theta term, three-gluon operator, and
EDMs and color EDMs of light quarks are important:
\be
 \de {\cal L} = && - \sum_{q=u,d,s} m_q\ov{q}(1+i\th_q\ga_5) q
               + \th_G \frac{\al_s}{8\pi}G\tilde{G} +wGG\tilde G\nonumber\\
 && -\frac{i}{2}\sum_{q=u,d,s}d_q \ov{q}F\si \ga_5 q
               -\frac{i}{2}\sum_{q=u,d,s}\tilde{d}_q \ov{q}g_sG\si \ga_5 q,
   \label{deL}
\ee
where $G\tilde{G}\equiv G^a_{\mu\nu}\tilde{G}^a_{\mu\nu}$,
$G\si \equiv t^aG^a_{\mu\nu}\si_{\mu\nu}$ and $GG\tilde{G}\equiv
f_{abc}G_{\mu\nu}^aG^b_{\nu\alpha}\tilde{G}^c_{\alpha\mu}$.
The coefficients in (\ref{deL}) are generated by the
CP violation in the SUSY breaking sector and evolved
down to 1 GeV, which is the border line of viability of the
perturbative quark-gluon description.

In this Letter we present a systematic study of the transition
from this Lagrangian to the effective T-odd meson-nucleon
interactions which determine the magnitude of the CP-violating
nuclear moments and the T-odd asymmetries in nucleon scattering.
T-odd nuclear forces are the main source for the EDMs of heavy
diamagnetic atoms (see e.q. \cite{KL}). The quality of constraints
imposed on supersymmetric models from a recently improved
measurement of the EDM of the xenon and mercury atoms
\cite{Michigan,
Seattle} (as well as of future experimental efforts
with the EDMs of diamagnetic atoms and the T-odd nucleon
scattering \cite{future}) depends crucially on the treatment of
QCD and nuclear effects, i.e. on the extraction of limits on
$\tilde d_i$ from the experimental bound on the atomic EDM. The
implications of this powerful constraint for the CP violation in
the supersymmetric models have been emphasized in Refs.
\cite{KZ,FOPR}, and numerically exploited in \cite{MadSuss}. The
purpose of this work is to give ``state-of-the-art'' estimates for
various T-odd nucleon-meson coupling constants, i.e to find their
best values in terms of the coefficients in eq. (\ref{deL}). This
problem is reminiscent of Ref. \cite{DDH} which estimates various
P-odd meson-nucleon couplings in terms of the Fermi constant.

T-odd nuclear forces inside the nucleus can be approximated by a meson exchange
with one of the meson-nucleon couplings being T-violating \cite{Barton}.
It is natural to expect that pion exchange dominates in the T-odd channel.
The coupling of nucleons with pions can be conveniently
parametrized \cite{Peter,HH} as
\begin{eqnarray}
 {\cal L}^{CP} = \bar g_{\pi NN}^{(0)}\bar N\tau^aN\pi^a
+\bar g_{\pi NN}^{(1)}\bar NN\pi^0
\end{eqnarray}
These couplings are generated by the theta term and
by the color EDMs of quarks, $\bar g \equiv \bar g (\bar\theta,~
\tilde d_i)$, where $\bar \theta = \theta_G + \sum \theta_q$.
Couplings that change the
isospin by two units can be generated only at the expense of
an additional $m_u-m_d$ suppression and are ignored in the present analysis.
$\gzero(\bar\theta)$ is
rather well known \cite{CDVW}, as it can be deduced from the size of the
$\langle N| \bar uu -\bar dd| N \rangle $ matrix element. For most of the
models of CP violation including minimal SUSY models, $\bar\theta$
has to be removed by the Peccei-Quinn (PQ) symmetry
leaving quark color EDMs as the dominant source for
CP-odd nuclear forces.
The contribution of the three-gluon operator $GG\tilde G$
to $\bar g$ is additionally suppressed by $m_q$ and
can be neglected.

The first step in the calculation of $\bar g_{\pi NN}^{(0)}(\bar\theta,~
\tilde d_u+ \tilde d_d)$ and $\bar g_{\pi NN}^{(1)}(
\tilde d_u - \tilde d_d)$ is the reduction of the pion field by means of
PCAC \cite{KKYZ}, Fig. 1a. The smallness of the $t$-channel pion momentum
compared to the characteristic hadronic scale justifies this procedure,
\be
\langle N \pi^a | {\cal O}_{CP} |N^\prime\rangle =
\fr{i}{f_\pi}\langle N|[ {\cal O}_{CP}, J^a_{05} ] |N^\prime\rangle.
\label{start}
\ee
The commutator of the zero component of the axial current with CP-violating
operators ${\cal O}_{CP}= \ov{q}g_sG\si \ga_5 q$ can be easily computed,
leading to the matrix elements of the $\ov{q}g_sG\si q$ operators over the
nucleon state \cite{KKYZ}. However, eq. (\ref{start}) is an incomplete
result.
A second class of contributions was pointed out in Ref.
\cite{FOPR} and in Refs.\cite{PR3,PR4} in the context of the neutron EDM
problem. They consist of the pion pole diagrams, Fig. 1b,
which contribute at the same order of chiral perturbation theory. Indeed,
the quantum numbers of the $\ov{q}g_sG\si q$ operators allow them to
produce zero-momentum $\pi^0$'s from the vacuum. The pion-nucleon scattering
amplitude at
vanishing pion momentum is proportional to the first power of the quark mass,
whereas the pion propagator contains $1/m_q$, so that Fig. 1b and 1a
both contribute at the same $O(\tilde d_qm_q^0)$ order.

\begin{figure}
 \centerline{%
   \psfig{file=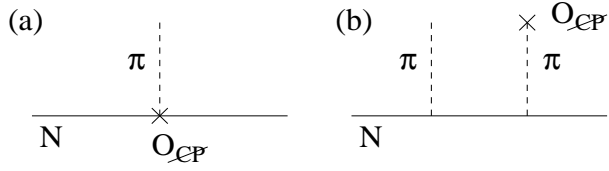,width=8cm,angle=0}%
         }
\vspace{0.1in}
 \caption{Two classes of diagrams contributing to the CP-odd pion-nucleon
coupling constant.}
\end{figure}

Using the low energy theorems that relate the pion-nucleon scattering amplitude
with the matrix elements of $m_q \bar q q $ over the nucleon state, we
arrive at the following intermediate result for the $NN\pi^0$ vertex:
\begin{eqnarray}
&&\fr{1}{2f_\pi}\langle N | \tilde d_u(\bar u g_sG\si u - m_0^2
\bar uu) - \tilde d_d (\dGd - m_0^2\dd )|N\rangle  \nonumber\\ &&+
\fr{m_*}{2f_\pi}\left[2\bar\theta + m_0^2\left(
\fr{\tilde d_u}{m_u}+\fr{\tilde d_d}{m_d}+\fr{\tilde d_s}{m_s}\right)
\right]\langle N | \uu-\dd|N\rangle.
\label{chir}
\end{eqnarray}
In this expression, $m_* = m_um_d/(m_u+m_d)$ and
$m_0^2 = \langle 0|\bar q g_s G\si q |0\rangle/\qq =-(0.8\pm 0.1)
$GeV$^2$ \cite{BK} parametrizes the strength of
the quark-gluon dim=5 vacuum condensate. In our
case, this originates from the
$\langle \pi_0 |\bar q g_s G\si\gamma_5 q |0\rangle$ matrix
element and the minus sign is included into the definition of $m_0^2$ for
convenience.
An alternative way of obtaining the amplitude (\ref{chir})
is to chirally rotate quark masses to the basis where
pions cannot be produced from the vacuum, $\langle \pi^0|
2\sum \ov{q}m_q\th_q\ga_5 q +
\sum \tilde{d}_q \ov{q}g_sG\si \ga_5 q|0\rangle = 0$, while keeping
the theta term fixed, $\sum \theta_q = {\rm const}$. This eliminates
diagrams 1b, but creates an additional contribution to 1a, leading to the
same result (\ref{chir}).

When the PQ mechanism is activated, removing the theta term, the minimum of
the axion vacuum is shifted from $\bar \theta = 0$
by the color EDM operators \cite{BUP}. It turns out that the
true minimum is such that the square bracket in eq. (\ref{chir}) is
zero so that only the first line survives.
This leads to a relatively simple expression for the couplings
\begin{eqnarray}
\gz=\fr{\tilde d_u+\tilde d_d}{2f_\pi}\langle p |H_u-H_d|p\rangle\\
\go=\fr{\tilde d_u-\tilde d_d}{2f_\pi}\langle N |H_u+H_d|N\rangle\nonumber
\end{eqnarray}
in terms of matrix elements of the $H_u$ and $H_d$ operators:
\begin{eqnarray}
H_u \equiv \uGu - m_0^2 \uu;\;
H_d \equiv \dGd- m_0^2 \dd
\end{eqnarray}
Previously, using a combination of QCD sum rules and scaling
arguments, Ref. \cite{KKYZ} estimated that $\langle N| \ov q g_s
G\si q   |N\rangle \sim \fr{5}{3}m_0^2\langle N| \ov q  q
|N\rangle$. Another analysis \cite{OSU} finds similar estimate
$\langle N| \ov q g_s G\si q   |N\rangle \sim m_0^2\langle N| \ov
q  q   |N\rangle$. Obviously, these estimates are not sufficient
to derive a reliable answer for $\gz$ and $\go$ because of the
additional $-m_0^2\langle N| \ov q  q   |N\rangle$ contribution
coming from diagrams 1b. The danger of mutual cancelation between
the two contributions was realized in Ref. \cite{FOPR} where the
need for a dedicated analysis of $ \langle N| H_{u(d)} |N\rangle$
was emphasized.
\begin{figure}
 \centerline{%
   \psfig{file=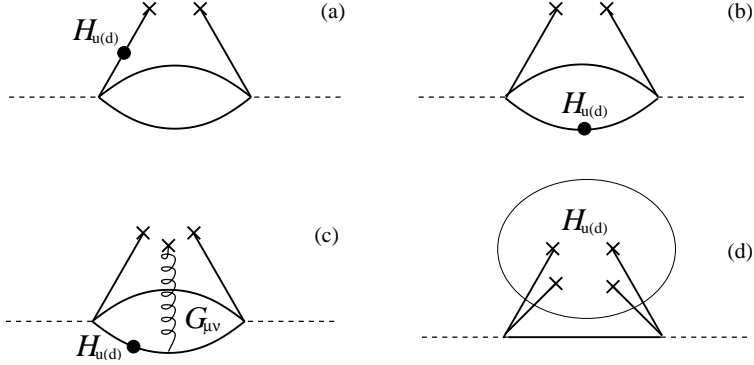,width=10cm,angle=0}%
         }
\vspace{0.1in}
 \caption{Various classes of diagrams contributing to the OPE for the
  $\psl$  structure.}
\end{figure}
In the rest of this paper we derive the QCD sum rules \cite{SR} for the
matrix elements of the $H_{u(d)}$ operators. The advantage of this approach is
that the operator product expansion (OPE) will contain similar vacuum
condensates for both sources, $\ov q g_s G\si q$ and $-m_0^2 \ov q  q $,
which allows us to trace possible cancelations. Following Refs. \cite{PR4,PRth},
we introduce the generalized nucleon interpolating current,
 \be
\eta_p & = & (j_1 +\beta j_2), \label{current}
\ee
which combines the two Lorentz structures,
$j_1 = 2\ep_{abc}(u_a^TC\ga_5d_b)u_c$ and
 $j_2 =  2\ep_{abc}(u_a^TCd_b)\ga_5u_c$.

We compute the OPE for the correlator of this
current in the presence of $\tilde d_{u(d)}
H_{u(d)}$ external sources
\be
 \Pi(Q^2) & = & i\int d^4x e^{ip\cdot x}
    \langle 0|T\{\eta_N(x)\ov{\eta}_N(0)\}|0\rangle_{\tilde d_u, \tilde d_d},
     \label{pi}
\ee
where $Q^2=-p^2$, with $p$ the current momentum.

We limit our calculation to the Lorentz structure proportional to
$\psl$ because it is less susceptible to direct instanton
contributions and excited resonances than the chirally even
structure proportional to {\bf 1}. Relevant diagrams for this
correlator are shown in Fig. 2. After a straightforward
calculation, we find
\be \label{OPE} \psl~ \Pi^{\rm OPE}(Q^2) =
\fr{\psl\qq}{\pi^2f_\pi}\left [ \ln\left( {\Lambda_{UV}^2\over
-p^2} \right ) \pi^{LO}+  \fr{1}{p^2}\ln\left(-p^2\over
{\Lambda_{IR}^2}\right)\pi^{NLO} + \fr{1}{p^2}\pi^{NNLO} \right ].
\ee
The leading order term is given by the diagrams 2a-2b,
\be
\pi^{LO} =
\fr{3m^2_0}{64}\left[d_-(5-2\beta-3\beta^2)+d_+(1-\beta)^2
\right]. \label{lo} \ee
Here we have introduced the combinations
$d_+ =\tilde d_u +\tilde d_d$ and $d_- =\tilde d_u -\tilde d_d$.
It turns out that the diagrams 2a, where the external source
enters through the $\langle 0|\bar q \slash{\!\!\!\!D} q|0\rangle$
structure,
 give large and opposite sign contributions for the
$g_s\bar q G\si q$ and $-m_0^2\bar qq$ sources so that their combined
effect in $H_q$ is nil. Fortunately, this cancelation does not hold for
diagrams 2b that give (\ref{lo}).

The next-to-leading term corresponds to diagrams 2c,
\be
\pi^{NLO}= \fr{3d_-}{32}\left[\fr{ m_0^4(1-\beta^2)-
g_s^2(GG)}{9}(7-2\beta-5\beta^2)\right]\nonumber\\
+\fr{d_+}{96}g_s^2(GG)(1-\beta)^2, \label{nlo} \ee
which
contribute to the OPE (\ref{OPE}) with the log of the infrared
cutoff $\Lambda_{IR}$. $g_s^2(GG)\simeq 0.4-1$ GeV$^4$ is the
vacuum gluon condensate. The next-to-next-to-leading order \be
\nonumber \pi^{NNLO}
= \fr{d_-}{24}\left[\chi_S\pi^2(1+2\beta -3\beta^2)\right.\\
\left.
+\fr{3m_0^4}{16}(13-2\beta-11\beta^2)\right] + \fr{d_+}{24}
\left[\chi_T\pi^2(1-\beta)^2\right.
\label{nnlo}
\\\nonumber\left.
+\fr{m_0^4}{16}(-11-14\beta+\beta^2)-\fr{g_s^2(GG)}{24}(5\beta^2+8\beta-1)
\right ].
\ee
contains the vacuum polarizabilities,
\be
\chi_{S,T} \equiv\!\int\!\! d^4x \langle 0|T\{\uu\pm\dd(x),
H_u\pm H_d(0)\}|0\rangle,
\label{chi}
\ee
and the vacuum factorization assumption has been made in (\ref{nnlo}).

The sum rules prescription involves matching the
OPE with the phenomenological part,
$\Pi^{{\rm OPE}}(Q^2)=\Pi^{\rm phen}(Q^2)$,
where
\be
\psl ~\Pi^{\rm phen} &=&\psl \left(
\fr{2\lambda^2\bar g_{\pi NN}m_N}{(p^2-m_N^2)^2} +
\fr{A}{p^2-m_N^2}\cdots\right)
          \label{phenfull}
\ee
contains double and single pole contributions, and the continuum.
After Borel transformation of the
sum rule $\Pi^{{\rm OPE}}(Q^2)=\Pi^{\rm phen}(Q^2)$ we obtain
\be
&&\fr{\qq}{\pi^2f_\pi}\left [\pi^{LO}E_0 -
\nonumber
\fr{\pi^{NLO}}{M^2}(\ln\left[M^2\over {\Lambda_{IR}^2}\right]-0.58)
- \fr{\pi^{NNLO}}{M ^2} \right ] =\\
&&M^{-4}\exp{\left[-\fr{m_N^2}{M^2}\right]}(2\la^2 m_N \bar g_{\pi
NN}+AM^2) +M^{-2}B\exp{\left[-\fr{s}{M^2}\right]} \label{main} \ee
Here $s$ is the continuum threshold and $E_0=1-{\rm e}^{-s/M^2}$.
$A$ and $B$ parametrize the contribution of excited states and are
assumed to be independent of $M$.

It is reasonable to start the numerical
treatment from a simple estimate, \'a la Ioffe \cite{Ioffe},
which assumes the dominance of the ground state and the
LO OPE term, and eliminates $\lambda$ using the nucleon mass sum rule
for $\psl$.
Separating different isospin structures, we find
\be
\gone = (\tilde d_u - \tilde d_d)~\fr{3}{2}~
\fr {4\pi^2|\qq|m_0^2}{m_Nf_\pi M^2}
~F_1(\beta)
\\
\gzero = (\tilde d_u +\tilde d_d)~\fr{3}{10}~
\fr {4\pi^2|\qq|m_0^2}{m_Nf_\pi M^2} ~ F_0(\beta).
\ee
Here $F_1(\beta)=
(5 - 2\beta - 3\beta^2)/(5 + 2\beta + 5\beta^2)$ and
$F_0(\beta)=5(1-\beta)^2/(5 + 2\beta + 5\beta^2)$.
$F_1(0)=$ $F_0(0)=1$.
To get numerical estimates, we choose $\beta = 0$,
extensively used in lattice simulations. It is well known
that the $j_1$ current has a much better overlap with the nucleon ground
state and $\lambda_1 \gg \lambda_2$.
Substituting $M = 1$GeV, we obtain
\be
\label{nnum1}
\gone = 3\times10^{-12}\fr{\tilde d_u- \tilde d_d}{10^{-26} {\rm cm}}
\fr{|\qq|}{(225 {\rm MeV})^3}~\fr{|m_0^2|}{0.8{\rm GeV}^2}
\\
\gzero =0.6\times10^{-12}\fr{\tilde d_u+ \tilde d_d}{10^{-26} {\rm cm}}
\fr{|\qq|}{(225 {\rm MeV})^3}~\fr{|m_0^2|}{0.8{\rm GeV}^2}
\label{nnum2}
\ee
In most SUSY models, $d_{u(d)} =$ loop factor $\times M_{SUSY}^{-2}
\times$ a linear combination of $m_u$ and $m_d$. When combined with
$\qq$ from eqs. (\ref{nnum1}-\ref{nnum2}), this forms $m_\pi^2f_\pi^2$ times
a function of $m_u/m_d$, thus eliminating a major source of uncertainty
in EDM calculations due to the poor knowledge of $m_u+m_d$
\cite{PR3,PR4,PRth}. The estimate (\ref{nnum1}) is twice smaller than the
value of $\gone$ used in \cite{FOPR}.  Also in agreement with \cite{FOPR},
eqs. (\ref{nnum1}-\ref{nnum2})
suggest that $\gzero/\gone \sim 0.2 d_+/d_-$.

For a more systematic analysis, one has to include $NLO$ and $NNLO$ terms in
the OPE. Here, we immediately face the problem of the
unknown vacuum condensates
$\chi_{S,T}$. Even though the vacuum
correlators $\langle \bar qq, \bar qq \rangle$ can be determined
using chiral perturbation theory \cite{Chir}, there is no direct
information on $\langle \bar qq, \bar qg_sG\si q \rangle$ other than
that it is likely to be
comparable with $m_0^2\langle \bar qq, \bar qq \rangle$.
At this point we would like to take advantage of the
possibility to choose $\beta$ in such a way as to minimize
higher order terms in the OPE. We note that $\chi_S$ in (\ref{nnlo})
is multiplied by $1+2\beta -3\beta^2$ which becomes 0 at $\beta = -1/3$ and 1.
The choice of $\beta = 1 $ also suppresses the leading order, while
$\beta = -1/3$ maximizes it.
For the expected size of $\chi_S$ \cite{Chir},
$\chi_S\sim\pm 0.16\times m_0^2\qq f_\pi^{-4}$, we can choose
$\beta$ in a range such that the whole square bracket in front of $d_-$
in eq. (\ref{nnlo}) is zero. This gives a range of interpolating currents
around $\beta = -1/3$,
\be
-0.5<\beta<0,
\label{beta}
\ee
where we can tune the $NNLO$ terms to zero in the $\gone$ channel.
Variation of $\beta$ in this range contributes to an estimate of the
uncertainty in our analysis.
In the  $\bar g_{\pi NN}^{(0)}$ channel there is no
obvious choice of $\beta$ that would remove $\chi_T$ and leave the leading
order term un-suppressed, so we will choose the same $\beta$ as for
$\bar g_{\pi NN}^{(1)}$.
We note that this range is close to $\beta=0$ as used for
(\ref{nnum1}-\ref{nnum2}).
One should also worry about the dependence on $\Lambda_{IR}$ in $NLO$.
Remarkably, in the range (\ref{beta}), this dependence
is softened by cancelation of the $m_0^4$ and $g_s^2(GG)$ terms.

The preferred range for $\gone$ and $\gzero$ is determined according to the
following procedure. We take the OPE side of (\ref{main}) at the lower
point of the usual Borel window, $M^2=0.8$ GeV$^2$, and vary it
through the range of parameters
$-0.5\le\beta\le0$,
$300{\rm MeV}\le\Lambda_{IR}\le500{\rm MeV}$,
$0.7 {\rm GeV}^2 \le |m^2_0| \le 0.9 {\rm GeV}^2$,
$0.4 {\rm GeV}^4 \le g_s^2(GG) \le 1 {\rm GeV}^4$, and
$2 {\rm GeV}^2 \le s \le 3{\rm GeV}^2$, $0.8{\rm GeV}^6\le
(2\pi)^4\lambda^2 \le 0.9 {\rm GeV}^6 $ and finally
$-6{\rm~GeV}^{-1}\le\chi_T/|\qq|\le6$
GeV$^{-1} $. On the r.h.s of (\ref{main}) we
assume the dominance of the double pole
contributions for $M^2=0.8$ GeV$^2$
and allow for a $50\%$ correction due to the presence of
the unknown parameters $A$ and $B$, thus effectively
widening the allowed range for $\bar g_{\pi NN}$. As expected,
the couplings are most sensitive to the
value of $m_0^2$. The final results are presented in Table 1.
Our ``best'' value for $\gone$ is determined by averaging over $\beta$ and
choosing the central values for condensates, which also
suppresses the logarithmic term. In order to separate the
contribution of $\gone$ from the $A$ and $B$ terms, we impose a relation
among $A$, $B$ and $\pi^{LO}$ obtained by requiring
the same large $M^2$ asymptotic
behavior for both sides of (\ref{main}). The resulting
sum rule is fitted numerically and produces a result 1.5 times smaller than
the naive estimate (\ref{nnum1}). For $\gzero$ the
best value cannot be determined
as the OPE side changes sign depending on the value of $\chi_T$.

Also included in this table are the preferred ranges for the CP-odd
couplings of nucleons with $\eta$, $\rho$ and $\omega$ mesons.
The couplings with $\rho$ and $\omega$ have the EDM-like structures
$-\fr{i}{2}\bar N (\partial_\nu V_\mu-\partial_\mu V_\nu) \sigma_{\mu\nu}
\gamma_5 N$ with properly arranged isospin indices.
They can be easily extracted
from the calculation of the
neutron EDM $d_n$, induced by $\tilde d_{u(d)}$ \cite{PR4} after a
simple re-assignment of charges for the external vector currents.
Best values  for $\bar g_{\rho NN}$ and $\bar g_{\omega NN}$ follow
from the central values of $d_n(\tilde d_u,\tilde d_d)$ given in \cite{PR4}.
Finally, the coupling to the $\eta$ meson is dominated by
the strange quark chromoelectric dipole moment $\tilde d_s$
in the isospin-singlet and by $\tilde d_u -\tilde d_d$ in the
isospin-triplet channels, and in both cases only the expected range can be
quoted.

\begin{table}
\begin{center}
\begin{tabular}{ccc}
    coupling  & preferred range & best value \\
\hline \hline

$ \bar g_{\pi NN}^{(1)}$ $\times 10^{12}$ & $(1$ to $6)~d_{-,26}$
&
 $2~d_{-,26}$
  \\
$\bar g_{\pi NN}^{(0)}$  $\times 10^{12}$ & $(-0.5$ to
$1.5)~d_{+,26}$
   &  $--$ \\

$\bar g_{\eta NN}^{(0)}$  $\times 10^{12}$
 &  $\sim (-1.5$ to $1.5)~d_{s,26}$  & $--$ \\

$\bar g_{\eta NN}^{(1)}$  $\times 10^{12}$
 &   $(-0.3$ to $1)~d_{-,26}$ & $-- $ \\

$\bar g_{\rho NN}^{(0)}$ & $(0.7$ to $2)~d_{+}$  &
 $1.4~d_{+}$
  \\
 $\bar g_{\rho NN}^{(1)}$ & $(0.5$ to $1.5)~d_{-}$  &
 $0.9~d_{-}$\\
$\bar g_{\omega NN}^{(0)}$  & $(-1.5$ to $-0.5)~d_{+}$  &
 $-0.9 ~d_{-}$\\
$\bar g_{\omega NN}^{(1)}$ &  $(-2$ to $-0.7)~d_{+}$  &
 $-1.4~d_{-}$
\\
\end{tabular}
\end{center}
\caption{Preferred ranges and best values for the CP-odd
meson-nucleon coupling constants induced by quark chromoelectric
dipole moments. All values contain an overall multiplier
$|\qq|/(225~{\rm MeV})^3$. $d_{-,26}$, $d_{+,26}$ and $d_{s,26}$
are $d_-$ and $d_+$ and $d_s$ in units of $10^{-26}$ cm.}
\label{table}
\end{table}

In conclusion, we have shown that the size of the CP-odd
pion-nucleon constant generated by quark chromoelectric dipole
moments is given by the matrix element of $\bar qg_s G\si q - m_0^2\bar qq$
over the nucleon state. We have constructed a QCD sum rule
for this matrix element and determined the preferred range and the
best value for the $\gone$ coupling. The upper part
of the preferred range agrees
with previous estimates. However, in the interpretation of the experimental
limit on the EDM of the mercury atom \cite{Seattle} in terms of limits
imposed on new CP-violating physics, a more conservative value
$\gone = 2 (\tilde d_u-\tilde d_d)/10^{-14}$cm should be used.
This translates the result of Ref. \cite{Seattle} (see \cite{KL,FOPR}
for details) into the bound $|\tilde d_u - \tilde d_d|< 2\times 10^{-26}$ cm.
This constraint provide a sensitive probe of CP violation in the
supersymmetric spectrum up to $M_{SUSY}$ of few TeV.

{\bf Acknowledgments}
The author would like to thank P. Herczeg, M. Voloshin and especially  A. Ritz
for helpful discussions.

\end{document}